# THE PRE-PERIHELION ACTIVITY OF DYNAMICALLY NEW COMET C/2013 A1 (SIDING SPRING) AND ITS CLOSE ENCOUNTER WITH MARS

Dennis Bodewits[1§], Michael S. P. Kelley[1], Jian-Yang Li[2], Tony L. Farnham[1], and Michael F. A'Hearn[1]

[1] Department of Astronomy, U. Maryland, College Park, MD 20742, USA. dennis@astro.umd.edu; msk@astro.umd.edu; farnham@astro.umd.edu; ma@astro.umd.edu. [§] Corresponding author

[2] Planetary Science Institute, 1700 East Fort Lowell Road, Suite 106, Tucson, AZ 85719, USA. jyli@psi.edu

## ABSTRACT

We used the UltraViolet-Optical Telescope on board Swift to systematically follow the dynamically new comet C/2013 A1 (Siding Spring) on its approach to the Sun. The comet was observed from a heliocentric distance of 4.5 AU pre-perihelion to its perihelion at 1.4 AU. From our observations, we estimate that the water production rate during closest approach to Mars was $1.5 \pm 0.3 \times 10^{28}$ molecules/s, that peak gas delivery rates where between 4.5 – 8.8 kg/s, and that in total between $3.1 – 5.4 \times 10^4$ kg cometary gas was delivered to the planet. Seasonal and evolutionary effects on the nucleus govern the pre-perihelion activity of comet Siding Spring. The sudden increase of its water production between 2.46 – 2.06 AU suggest the onset of the sublimation of icy grains in the coma, likely driven by $CO_2$. As the comet got closer to the Sun, the relative contribution of the nucleus' water production increased, while $CO_2$ production rates decreased. The changes in the comet's activity can be explained by a depletion of $CO_2$, but the comet's high mass loss rate suggests they may reflect primordial heterogeneities in the nucleus.

3 figures, 2 tables

## 1. INTRODUCTION

Comets formed early in the evolution of the solar system while material was accreting to form planets. When proto-planets became large enough, a population of comets was dynamically ejected into the Oort cloud. Comets entering the inner solar system for the first time are called dynamically new comets. These objects retain some of the most primordial material available for observation in the solar system. Through dynamical evolution, the orbits of Oort Cloud comets can evolve to shorter periods. Halley-type comets are found on the extreme end of this orbital evolution (Oort 1950; Levison 1996).

Dynamically new comets have not been significantly heated by the Sun since they were placed into the Oort Cloud. On approach, their brightness might increase rapidly at large heliocentric distance, and then slower near perihelion (Oort & Schmidt 1951; Whipple 1978). The enhanced activity may be attributed to the evaporation of a crust created by interstellar radiation (Johnson et al. 1987; Whipple 1978; Stern 2003). Volatiles such as CO and $CO_2$ may drive off icy grains, which have relatively high albedos, increasing the amount of sunlight reflected by the coma (A'Hearn et al. 1995). Outbound, new comets behave more like evolved Oort Cloud comets. It is unknown in how far such behavior is driven by comet evolution (the removal of outer layers) or by primordial heterogeneity reflecting comets' origins within the solar nebula.

Few dynamically new comets have been systematically studied at large heliocentric distance and throughout their orbit (cf. A'Hearn et al. 1984; Meech et al. 2009). Here, we present the results of our long-term monitoring campaign of comet C/2013 A1 (Siding Spring) using the Swift space telescope facility. The comet was first discovered 7.2 AU from the Sun, and passed Mars at a distance of 141,000 km on Oct. 19, 2014. Calculations of its original reciprocal semi-major axis suggest that Siding Spring is a dynamically new comet (Nakano 2014; Williams 2014).

## 2. OBSERVATIONS AND DATA ANALYSIS

Swift is a multi-wavelength observatory equipped for rapid follow-up of gamma-ray bursts (Gehrels et al. 2004). Our observations use its Ultraviolet and Optical Telescope (UVOT) that provides a 17 × 17 arcminute field of view, a plate scale of 1 arcsec/pixel, and a point spread function of 2.5" FWHM (Mason et al. 2004). Swift/UVOT is equipped with 7 broadband filters that cover wavelengths between 1800 and 6000 Å. The detector is a microchannel plate intensified CCD, and each incident photon on its photocathode is amplified million-fold using a three stage multichannel plate.

Swift observed C/2013 A1 (Siding Spring) 10 times between November 2013 and October 2014 (Table 1). Each observation consisted of at least two orbits, and during each orbit several exposures were acquired with the UVW1 and V-band filters. The U-band filter was used only during the two visits in October 2014. Swift tracked the sky at the sidereal rate and used short, 200-s exposures to minimize smearing due the comet's motion. The comet's apparent motion warrants different sky backgrounds between visits, and background stars are suppressed effectively by stacking individual exposures using a resistant mean algorithm (Fig. 1). This algorithm is less effective in the V-band, where fewer individual exposures are available.

The broadband filters provide a measure of the comets' water and dust production rates (Bodewits et al. 2014). The UVW1 filter (central wavelength 2600 Å, FWHM 700 Å) is well placed to observe the three very strong OH vibrational transitions between 2811 and 3122 Å, and the U filter (central wavelength 3501 Å, FWHM 875 Å) contains the strong CN Δv = 0 band at 3876 Å. Both filters also contain reflected continuum. To remove this, we weigh a solar spectrum with the UVOT filter transmissions to determine how much the continuum contributes to the UVW1 and U flux. Li et al. (2014) measured a reddening of 5% per 1000 Å between 4380 Å and 5890 Å in October 2013, increasing to 9% per 1000 Å in March 2014, within a 5000 km radius aperture. We calculated the ratio between the fluxes measured in the two filters for both reddening slopes, which gives the continuum removal factors. Those are UVW1/V = 0.135 – 0.130, and U/V = 0.508 – 0.484 for 5 and 9% per 1000 Å, respectively (for units of flux). We then measured fluxes in circular apertures with radii between 25 – 150 arcseconds.

Correcting for the filter transmission at the relevant wavelengths, the measured fluxes can be converted into column densities using heliocentric distance and velocity dependent fluorescent efficiencies (Schleicher & A'Hearn 1988). To derive water production rates, we compare the measured OH content of the coma with an OH distribution calculated using the vectorial model (Festou 1981; Combi et al. 2004) and assuming that the bulk outflow velocity decreases with the heliocentric distance as $v = 0.86\ r^{-1/2}$ km/s (Delsemme 1982).

At magnitudes between $15.5 < m_V < 10$ and $18 < m_{UVW1} < 11$ in the apertures used for this analysis (Table 1), Siding Spring was easily observable by Swift. UVOT is well calibrated, and systematic errors are of order 5% (Poole et al. 2008). Its filters are broadband filters and the continuum removal adds significantly to the uncertainty of our water production rates. The continuum contribution to flux in the UVW1 filter decreased from 88% on $r_h = 2.46$ AU to 21% on $r_h = 1.5$ AU, after which it increased again. The effect of the assumed reddening depends on this continuum contribution; increasing the reddening from 5 to 10% per 1000 Å results in a 40% larger water production rate at $r_h = 2.46$ AU, and a 9% higher production rate at 1.40 AU. The effect is <5% for all other measurements. Similarly, $C_2$ may contribute as much as 20% to the flux in the V-band (Bodewits et al. 2014), which can lead us to underestimate water production rates. The effect on water production rates is less than 10% between $r_h = 1.7 – 1.41$ AU, but may be as large as 20 – 40% for the observations at $r_h = 2.05$ and 1.40 AU, respectively. However, given the consistency between the $r_h = 1.40$ AU and the better-constrained measurement on $r_h = 1.41$ (<7%), we expect the effect of contamination on these observations to be minor.

### 3. RESULTS

*3.1 Water Production Rates and Active Area*

Water production rates are summarized in Fig. 2. Between 4.5 and 3 AU, we derived 3-sigma upper limits of $10^{27}$ molecules/s from the UVW1 images, consistent with the production rate derived from the first positive detection of OH emission at a heliocentric distance of 2.46 AU. The production rate then increased very rapidly from $2.42 \pm 0.9 \times$ $10^{27}$ molecules/s at 2.46 AU to $1.11 \pm 0.1 \times 10^{28}$ molecules/s at 2.1 AU, a 5-fold increase, but reached a plateau between 2.0 AU and perihelion (Oct. 25.3, 2014 at $r_h$ = 1.399 AU).

To further investigate the behavior of the comet's water production rate, we derived the required minimum active area corresponding to the measured water production using the sublimation model by (Cowan & A'Hearn 1979), assuming that every surface element has constant solar elevation (as would be the case if the rotational pole were pointed at the sun or if the nucleus was very slowly rotating) and is therefore in local, instantaneous equilibrium with sunlight. This maximizes the sublimation averaged over the entire surface, and results in a minimum total active area. We further assumed a Bond albedo of 0.02 (see e.g. Li et al. 2013) and 100% infrared emissivity. The results are shown in Fig. 3. The computed active area of Siding Spring is highly variable over time. The steep increase in the water production rate between 2.46 and 2.1 AU requires a 4-fold increase in the active area, from 4.4 to 17 km². The active area then decreased to ~9 km² around perihelion. It is also interesting to note that the comet's active area within 2.1 AU from the Sun was larger than those derived from upper limits for the observations of the comet at 3.49 and 3.23 AU (14 and 8.4 km², respectively). If the activity were driven by sublimation from a constant area on the nucleus, Swift would have detected OH much earlier. This disparity thus suggests that the origin of the comet's water changed between 3 and 2 AU pre-perihelion.

### *3.2 Comet Gases Delivered to Mars*

Before the close encounter, it was predicted that the upper atmosphere of Mars might be severely perturbed by the mass and energy delivered by neutral or ionized cometary gases (Yelle et al. 2014; Gronoff et al. 2014). Interpolating the values measured on October 13 and 23, 2014, we estimate the water production rate during the comet's close encounter with Mars on October 19 to be $1.5 \pm 0.3 \times 10^{28}$ molecules/s, where the range represents the difference between the production rates measured on 13 and 23 October, 2014. The last infrared observations of the comet occurred a month before encounter, on Sep. 21, 2014, and NEOWISE observations by Stevenson et al. (2015) yield a $CO_2$ production rate of $1.18 \pm 0.12 \times 10^{27}$ molecules/s. However, the NEOWISE observations cannot directly discriminate between $CO_2$ and CO, as both molecules have an emission band in the 4.6 μm filter. The fluorescence efficiency of the $CO_2$ v3 band is 11 times stronger than the CO 1 → 0 band (Crovisier & Encrenaz 1983). Given that the $CO/CO_2$ production rate ratio is typically of order 1, $CO_2$ is most likely responsible for the NEOWISE observed gas emission.

Assuming a constant $CO_2$ production rate of $1.2 \times 10^{27}$ molecules/s, we calculated column densities and particle fluxes for the comet gas delivered to Mars using a Haser model (Table 2; see Bodewits et al. 2011 for details). Most of the comet's water has been photodissociated into OH, the dominant water fragment at 141,000 km. At this distance from the nucleus, OH is the dominant molecule, both in number and in mass, followed by $H_2O$ and atomic oxygen. Peak gas mass delivery rates to Mars reached 4.5 kg/s during closest approach and totaled $3.0 \times 10^4$ kg. If CO is instead the molecule responsible for the IR flux, with a production rate of $1.3 \times 10^{27}$ molecules/s, it would have been the dominant molecule delivered to Mars. In that scenario, a total mass of $5.4 \times 10^4$ kg would have been delivered to Mars, with a peak flux of 8.8 kg/s, i.e. almost twice as much as in the $CO_2$- scenario. Based on HST observations from early 2014 (Li et al. 2014), dynamical models showed that only dust grains that can reach the Martian system are those ejected with speeds of a few meters per second, have radii of 0.7–3.6 mm, and are ejected at least 1.5 yr ($r_h$ > 3 AU) prior to the encounter (Kelley et al. 2014; Tricarico et al. 2014). The activity of the comet during closest approach to Mars does not affect the predicted dust delivery of 100 kg (Kelley et al. 2014). The mass of the gas delivered to Mars thus exceeds the mass of the dust by two orders of magnitude.

## 4. DISCUSSION

The activity of comet nuclei is complex. They may exhibit varying amounts of activity across their surfaces due to heterogeneities in the nucleus, local topography, due to physical evolution, etc. In this section, we will discuss what processes may have driven the changes in activity of Comet Siding Spring.

After $H_2O$, $CO_2$ and CO are the most abundant ices in comet nuclei and both gases can drive comet activity (A'Hearn et al. 2011; Ootsubo et al. 2012; Feaga et al. 2014). For Siding Spring, $CO_2$ production rates were reported by Stevenson et al. (2014) and Kelley et al. (2014). Between 3.8 and 3.1 AU, they report $CO_2$ production rates of $3.5 \times 10^{26}$ molecules/s, which, compared to our upper limit of $Q(H_2O) < 10^{27}$ molecules/s, suggests a $CO_2/H_2O$ abundance of at least 35% at those heliocentric distances, consistent with other comets observed at 4 to 3 AU by Ootsubo et al. (2012). $CO_2$ production rates initially increased as the comet reached a heliocentric distance of 1.9 AU but then decreased as the comet moved further inwards. Comparing this again to the water production rates, the $CO_2/H_2O$ coma abundance decreased from 13% to 9% between 1.9 and 1.5 AU. This decrease within 2 AU is more pronounced when calculating the active area required for the $CO_2$ production (Fig. 3), which appears to be more or less constant outside 1.9 AU. While we recognize the low-number statistics here, the variation of the $CO_2$ area (20%) might be a rotational effect and is consistent with the amplitude of the variation in the water production rates measured on Oct. 13 and Oct. 23.

The water production and corresponding active area increases steeply between 2.46 and 2.1 AU pre-perihelion, whereas the active area of $CO_2$ appears to be constant during that period. Swift's non-detection of OH between 3 and 4 AU at levels where the comet should have been detected had the comet had a constant active area on approach, suggests that's an additional source of water was activated between 2.46 and 2.06 AU. Based on the varying orientation of jets projected in sky plane near the nucleus, Li et al. (2014) presented two possible orientations for the rotational pole of the nucleus. In the first scenario, the subsolar point would move quickly from the one hemisphere to the other between 2.0 AU and perihelion, crossing the equator in early September, 2014 ($r_h \sim 1.55$ AU). In their second solution, the subsolar point remains on one hemisphere until $r_h \sim 1.5$ AU post-perihelion. Both pole solutions predict a very small change in latitude for the subsolar point on the comet's surface (5-10 degrees) at >2 AU inbound, making it unlikely that this change in activity reflects a heterogeneous distribution of ices. Models suggest that isothermal sublimation rates increase by two orders of magnitude between 3 – 1.6 AU (Cowan & A'Hearn 1979). We therefore attribute the sudden increase in the water production rate between 2.46 and 2.06 AU pre-perihelion to the increased sublimation rate of icy grains. The presence of icy grains is consistent with the spatial and temporal variations of the color in the dust coma within 5,000 km from the nucleus as observed by HST at 4.6, 3.8, and 3.3 AU (Li et al. 2014).

Water production rates plateaued between 1.7 and 1.4 AU, and the active areas of both $H_2O$ and $CO_2$ decreased. Sublimation rates of the nucleus' active areas likely increased with the increased solar flux, increasing the relative contribution of the nucleus to the water production (this increased efficiency is especially important when assuming that a large fraction of the nucleus was active). In both pole solutions by Li et al. (2014) the latitude of the sub-solar point changed by 50 – 60 degrees during this period. In one solution new parts of the surface may have become active; in the other solution, the sub solar point remained in at latitudes that had been heated before. The decrease of the $CO_2$ abundance may thus be due to a paucity of $CO_2$ on the hemisphere illuminated near perihelion, or due to evolutionary processes, such as the retreat of the more volatile $CO_2$ ice.

The behavior of C/2013 A1 (Siding Spring) is strikingly similar to that of comet C/2009 P1 (Garradd) before its perihelion, a bright Oort Cloud comet that could be observed throughout most of its apparition. Garradd's activity was highly asymmetric around perihelion (Bodewits et al. 2014; Feaga et al. 2014). Inbound, observations suggested a strong extended source of water (Paganini et al. 2012; Combi et al. 2013; Bockelée-Morvan et al. 2012), that were not removed during the comet's distant outburst but instead were continuously replenished (Combi et al. 2013). By comparing water production rates derived from Swift/UVOT observations with those acquired with instruments using smaller apertures, Bodewits et al. were able to separate the two sources of water. They concluded that between $r_h$ = 3.0 AU and perihelion ($r_h$ = 1.54 AU), $H_2O$ was produced predominantly from ice in the coma (Bodewits et al. 2014). The disappearance of the grains coincided with a strong decrease in the ratio between $CO_2$ and $H_2O$ production rates between 3 and 2 AU, from which we tentatively concluded that a disappearing $CO_2$ source drove the icy grains into the coma of comet Garradd (Bodewits et al. 2014; Decock et al. 2013; McKay et al. 2015).

It has been suggested that dynamically new comets are more active than short period comets (Meech et al. 2004). The minimum active area of 4.4 km$^2$, derived from the production rates at 2.46 AU, is likely the closest approximation to the contribution of the nucleus' surface to the total water production rate. If this is indeed representative of the size of the nucleus it implies a minimum radius of 600 m, consistent with current size estimates from MRO/HiRISE (Tamppari et al. 2014). Integrating the water and $CO_2$ production rates between May 28, 2014 and Oct. 23, 2014, we find that the comet lost 1.6 × 10$^{35}$ $H_2O$ molecules and 2 × 10$^{34}$ $CO_2$ molecules, equivalent to 6 × 10$^9$ kg of ice. Assuming a dust-to-gas ratio of 4 as was observed by Rosetta around 67P/Churyumov-Gerasimenko (Rotundi et al. 2015), the comet lost approximately 3 × 10$^{10}$ kg of material on its way to perihelion. If the comet indeed were only 600 m in radius, this would be as much as 10% of its entire mass, or a global layer of 20 m. Such as a large mass loss rate would imply a continuous exposure of fresh material on the comet's approach to perihelion, which would suggest that the observed changes in activity reflected a primordial heterogeneity of the nucleus.

## 5. SUMMARY

We obtained optical and UV images of comet C/2013 A1 (Siding Spring) using the Swift/UVOT from 4.5 – 1.4 AU pre-perihelion. The comet's water production rate did not follow a simple relation with increasing insolation but changed over time. The comet's activity may be summarized as follows:

- $r_h$ > 2.5 AU: $CO_2$ sublimation from a constant area on the nucleus, $H_2O$ production too low to detect.
- 2.5 < $r_h$ < 2.0 AU: Increased sublimation rate of icy grains rapidly increases the water production.
- $r_h$ < 2.0 AU: Rapid change of the subsolar latitude, resulting in insolation of new parts of the surface depending on pole solution. Effective water sublimation from nucleus begins, and $CO_2$ production rate decreases.

The changes in the comet's activity may be explained by seasonal and evolutionary processes, and the comet may have shed as much as 10% of its mass on approach. To disentangle the effect of these processes it is necessary to compare the activity behavior of a large sample of dynamically new comets to more evolved Oort Cloud comets.

We thank the Swift team for use of observing time granted through the Director's Discretionary program and the Guest Investigator program, and for the careful and successful planning of our observations. DB and TLF received support through the Swift Guest Investigator program (Swift Cycle 10, program NNH13ZDA001N). This research was supported by a contract to the University of Maryland by the NASA JPL Mars Critical Data Products Program.

## REFERENCES

A'Hearn, M. F., Schleicher, D. G., Millis, R. L., Feldman, P. D., & Thompson, D. T. 1984, AJ, 89, 579
A'Hearn, M. F., Millis, R. C., Schleicher, D. O., Osip, D. J., & Birch, P. V. 1995, Icar, 118, 223
A'Hearn, M.F. et al., 2011. Sci, 332(6), 1396
Biver, N., Bockelée-Morvan, D., Colom, P., et al., 1997. Sci, 275, 1915
Bockelée-Morvan, D., Biver, N., Swinyard, B., et al. 2012, A&A, 544, L15
Bodewits, D., Farnham, T. L., A'Hearn, M. F. A., et al., 2014. AJ, 786(1), 48
Combi, M. R., Mäkinen, J. T. T., Bertaux, J.-L., et al. 2013, Icar, 225, 740
Combi, M. R., Harris, W. M., & Smyth, W. H. 2004, in Comets II, ed. M. Festou, H. U. Keller, & H. A. Weaver (Tuscon, AZ: Univ. Arizona Press), 523
Cowan, J.J. & A'Hearn, M.F., 1979. M&P, 21, 155
Crovisier, J., & Encrenaz, Th., 1983, A&A 126, 1, 170
Decock, A., Jehin, E., Hutsemekers, D., & Manfroid, J. 2013, A&A, 555, 34
Delsemme, A.H., 1982. In: Comets. (Tucson, AZ: Univ. Arizona Press), 85
Feaga, L. M., A'Hearn, M. F., Farnham, T. L., et al. 2014, AJ, 147, 24
Festou, M.C., 1981. A&A, 95, 69
Gehrels, N., Chincarini, G., Giommi, P., et al. 2004, ApJ, 611, 1005
Gronoff, G., Rahmati, A., Wedlund, C. S., et al., 2014. GRL 41(1), 4844

Huebner, W. F., Keady, J. J., & Lyon, S. P., 1982, ASS, 195, 1
Johnson, R. E., Cooper, J. F., Lanzerotti, L. J., & Strazzula, G. 1987, A&A, 187, 889
Kelley, M. S. P., Farnham, T. L., Bodewits, D., et al., 2014. ApJ, 792(1), L16
Li, J.-Y., Besse, S., A'Hearn, M.F., et al., 2013. *Icar.*, 222(2), pp.559
Li, J.-Y., Samarasinha, N. H., Kelley, M. S. P., et al., 2014. ApJ, 797(1), L8
Levison, H.F., 1996. In" *Completing the Inventory of the Solar System*, ASP Conference Series, Eds. T.W. Rettig & J. Hahn,107, 173.
Mason, K. O., Breeveld, A., Hunsberger, S. D., et al. 2004, Proc. SPIE, 5165, 277
Meech, K. J., Hainaut, O. R., & Marsden, B. G. 2004, Icar, 170, 463
Meech, K. J., Pittichová, J., Bar-Nun, A., et al. 2009, Icar, 201, 719
McKay, A. J., Cochran, A. L., DiSanti, M. A., et al. 2015, Icar, 250, 504
Nakano, S., 2014. NK 2757, http://www.oaa.gr.jp/~oaacs/nk/nk2757.htm
Oort, J.H. 1950 *Bull. Astr. Inst. Ned.*, 11, 91
Oort, J.H. & Schmidt, M., 1951. BAN, 11, 259
Ootsubo, T., Kawakita, H., Hamada, S., et al., 2012. ApJ, 752(1), 15
Paganini, L., Mumma, M. J., Villanueva, G. L., et al. 2012, ApJL, 748, L13
Poole, T. S., Breeveld, A. A., Page, M. J., et al. 2008, MNRAS, 383, 627
Protopapa, S., Sunshine, J., Feaga, L., et al., 2014. Icar, 238, 191
Rotundi, A., Sierks, H., Della Corte, V., et al., 2015, Science, in press
Schleicher, D.G. & A'Hearn, M.F., 1988. ApJ, 331, 1058
Stern, S.A., 2003. Natur, 424, 639
Stevenson, R., Bauer, J. M., Cutri, R. M., et al., 2014. arXiv:142.2117
Tamppari, L., Zurek, R., Cantor, B., et al., American Geophysical Union Fall Meeting 2014, P42A-02
Tricarico, P., Samarasinha, N., Sykes, M. et al., 2014, ApJ, 787, 2, L35
Whipple, F.L., 1978. M&P, 18, 343
Williams, G. V., 2014, Observations and orbits of comets, MPEC 2014-Y44
Yelle, R.V., Mahieux, A., Morrison, S. et al., 2014. Icar, 237, 202

**Table 1:** Summary of the observing log. $\rho$ is the radius of the aperture at the distance of the comet used to derive water production rates $Q$. The errors and upper limits are 3-$\sigma$ stochastic errors, and for the magnitudes systematic errors are given. For the first four observations (4.54 – 3.23 AU) we assumed a reddening of 5% per 1000 Å, after that 10% per 1000 Å for the continuum removal.

| Date (UT) | $r_h$ (AU) | D (AU) | $\rho$ (km) | Q $H_2O$ ($10^{27}$ molec/s) | m | m |
|---|---|---|---|---|---|---|
| 11/02/2013 | 4.54 | 3.98 | 7.22E+04 | < 5.5 | 15.5±0.01 | 17.8±0.03 |
| 12/28/2013 | 4.01 | 3.63 | 5.26E+04 | < 2.8 | 15.4±0.01 | 17.9±0.03 |
| 02/17/2014 | 3.49 | 3.78 | 5.48E+04 | < 3.2 | 15.2±0.01 | 17.9±0.03 |
| 03/16/2014 | 3.23 | 3.79 | 5.49E+04 | < 2.9 | 15.1±0.01 | 17.6±0.03 |
| 05/29/2014 | 2.46 | 3.00 | 5.45E+04 | 2.42 ± 0.92 | 14.1±0.01 | 16.1±0.03 |
| 07/09/2014 | 2.06 | 2.07 | 1.12E+05 | 11.1 ± 1.0 | 12.0±0.01 | 13.4±0.03 |
| 08/19/2014 | 1.69 | 1.04 | 1.13E+05 | 12.1 ± 0.27 | 10.3±0.01 | 11.2±0.03 |
| 09/18/2014 | 1.50 | 0.99 | 1.08E+05 | 13.0 ± 0.45 | 10.6±0.01 | 11.1±0.03 |
| 10/13/2014 | 1.41 | 1.48 | 8.05E+04 | 12.5 ± 0.62 | 11.7±0.01 | 12.4±0.03 |
| 10/23/2014 | 1.40 | 1.70 | 9.25E+04 | 17.2 ± 0.50 | 9.8±0.01 | 11.6±0.03 |

**Table 2.** Total comet gas column densities, and peak fluxes, and total gas mass at Mars. Only photodissociation of $H_2O$ is considered for the production of fragment species.

| Gas | Production Rate ($10^{27}$ molec. s$^{-1}$) | Rel. Abundance (%) | N ($10^{15}$ m$^2$) | $F_{peak}$ ($10^{11}$ m$^{-2}$ s$^{-1}$) | Total Mass (kg) |
|---|---|---|---|---|---|
| $H_2O$ | 15 | 100 | 6.3 | 14.7 | 6,885 |
| OH | - | - | 10.3 | 15.2 | 10,543 |
| O | - | - | 7.3 | 6.2 | 7,002 |
| H | - | - | 3.5 | 3.2 | 209 |
| $CO_2$ | 1.18 | 0.08 | 1.86 | 3.0 | 4,929 |
| **TOTAL** | **16.2** | | **30.6** | **44.5** | **29,569** |

**Figure 1**: Swift/UVOT observations of C/2013 A1 on Oct. 13, 2014. Each image consists of several median stacked exposures to improve SNR and remove background stars. All images have the same physical scale (430 arcsec, equivalent to 440,000 km at the comet), are oriented in sky-coordinates (north up, east to the left), and have been individually stretched logarithmically for optimal presentation. Left: UVW1 - V (OH). Middle: V-band. Right: U – V (CN). Black crosses mark the position of the nucleus. White circles indicate a distance of 141,000 km, the separation between the comet and Mars during closest approach.

**Figure 2:** Water production rates of C/2013 A1 as function of heliocentric distance. Downward arrows indicate 3-sigma upper limits. Filled, red triangles indicate $CO_2$ production rates from Stevenson et al. (2014) and Kelly et al. (2014). Most error bars are the size of the symbols.

**Figure 3:** Active area for $H_2O$ (black circles) and $CO_2$ (red squares), calculated from measured production rates and assuming a local thermal equilibrium model. Most error bars are the size of the symbols.

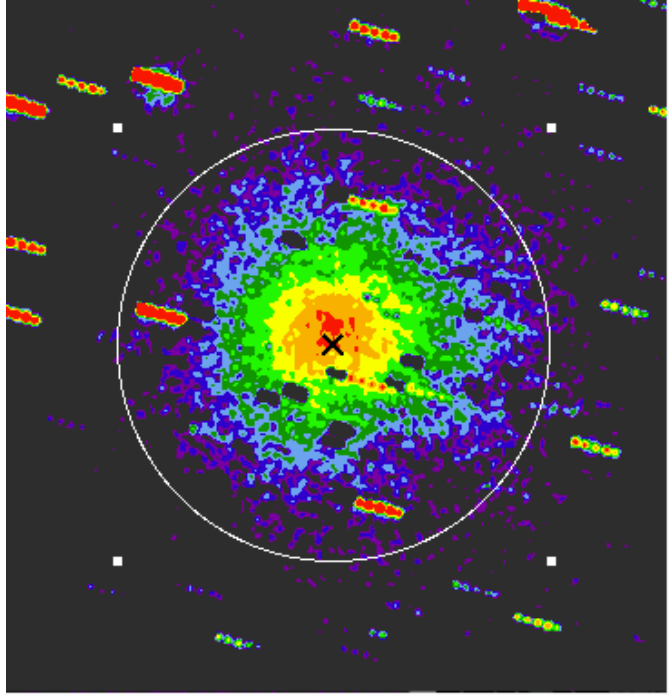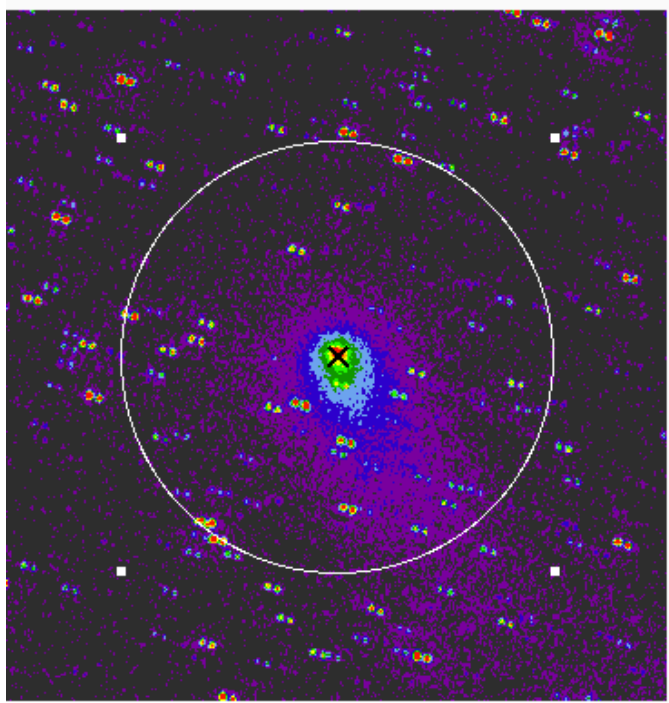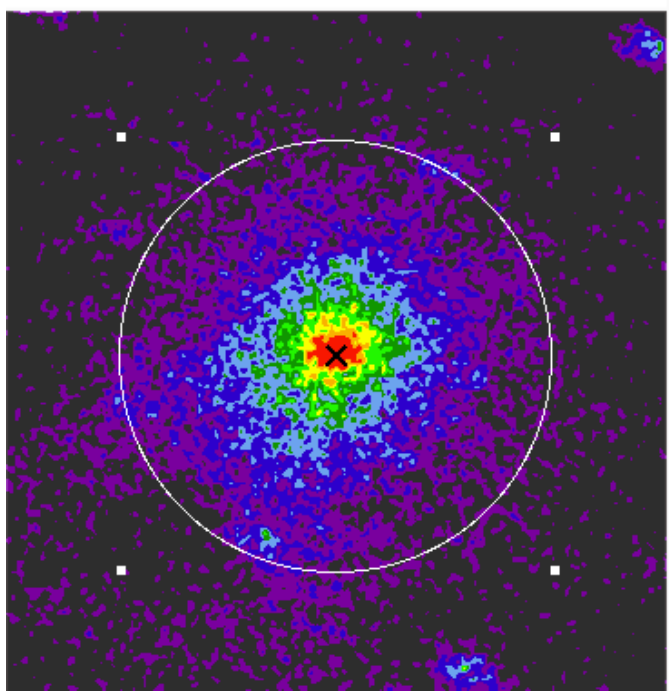

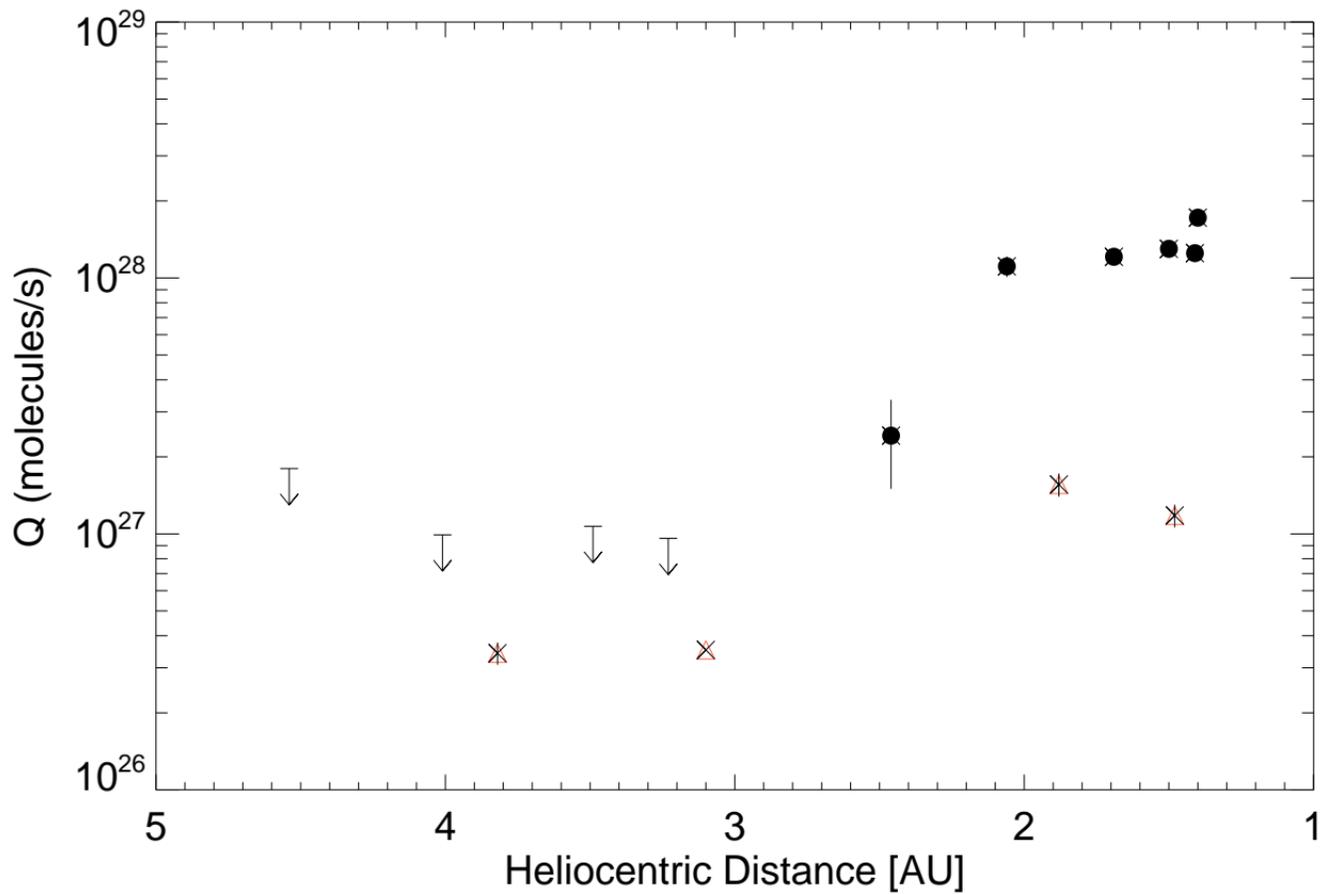

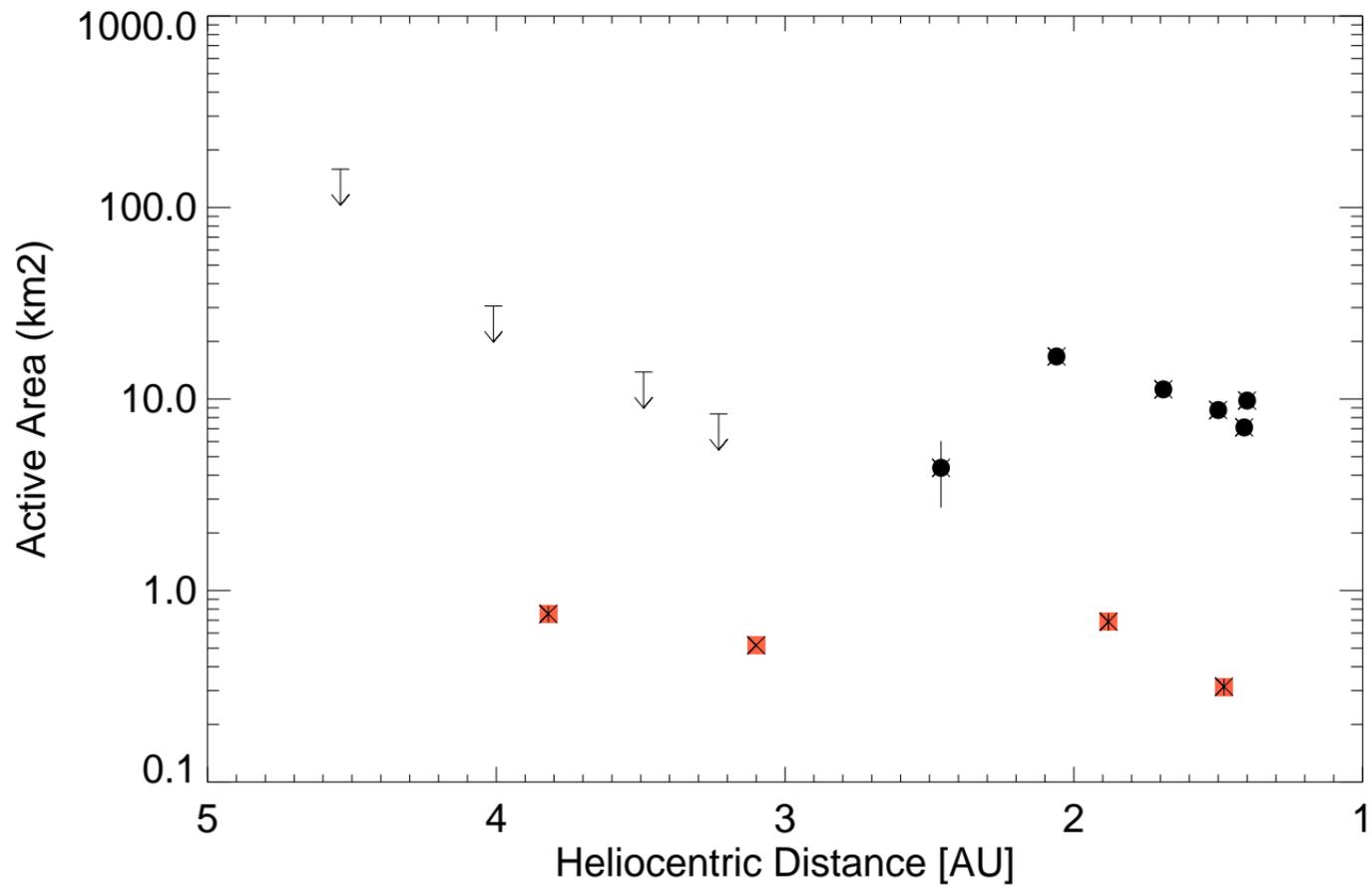